\documentclass[prd,twocolumn,floatfix,amsmath,nofootinbib,amssymb,floatfix]{revtex4}
\usepackage{graphicx,color,dcolumn,booktabs,bm}
\usepackage{longtable,lscape}
\usepackage{pdfpages}
\usepackage{txfonts}
\usepackage{overpic}
\usepackage{amssymb}
\usepackage{makecell}
\usepackage{indentfirst}
\usepackage{feynmf}   
\usepackage{slashed}  
\usepackage{cases}
\usepackage{color}
\usepackage{multirow}
\usepackage{makecell}
\usepackage{threeparttable}
\usepackage{epstopdf}
\usepackage{enumerate}
\usepackage{diagbox}
\usepackage{graphicx,color,dcolumn,booktabs,bm}
\usepackage[colorlinks,
            citecolor=blue,
            anchorcolor=red,
            menucolor=red,
            linkcolor=red,
            filecolor=red,
            runcolor=red,
            urlcolor=blue,
            frenchlinks=red]{hyperref}
\usepackage{tikz}

\begin{document}

\title{Scalar resonance contributions in the $D_{s1}(2460)^{+} \rightarrow D_{s}^{+}\pi^{+}\pi^{-}$ reaction}

\author{Zhong-Yu Wang$^{1}$}
\email{zhongyuwang@gzu.edu.cn}

\author{Yu-Shuai Li$^{2}$}
\email{liysh@pku.edu.cn}

\author{Si-Qiang Luo$^{3,4,5,6,7}$}
\email{luosq15@lzu.edu.cn}

\affiliation{
$^1$College of Physics, Guizhou University, Guiyang 550025, China \\
$^2$School of Physics and Center of High Energy Physics, Peking University, Beijing 100871, China\\
$^3$School of Physical Science and Technology, Lanzhou University, Lanzhou 730000, China\\
$^4$Lanzhou Center for Theoretical Physics, Key Laboratory of Theoretical Physics of Gansu Province, Lanzhou University, Lanzhou 730000, China\\
$^5$Key Laboratory of Quantum Theory and Applications of MoE, Lanzhou University,
Lanzhou 730000, China\\
$^6$MoE Frontiers Science Center for Rare Isotopes, Lanzhou University, Lanzhou 730000, China\\
$^7$Research Center for Hadron and CSR Physics, Lanzhou University and Institute of Modern Physics of CAS, Lanzhou 730000, China
}

\date{\today}

\begin{abstract}

Inspired by the newly observed $T_{c\bar{s}}$ state in the $D_{s1}(2460)^{+} \rightarrow D_{s}^{+}\pi^{+}\pi^{-}$ reaction by the LHCb Collaboration, we investigate the amplitude of this decay to explore the origins and properties of the open-charm tetraquark state based on the final state interaction. 
The invariant mass distributions of $D_{s}^{+}\pi^{+}$ and $\pi^{+}\pi^{-}$ are well reproduced by the $S$-wave scattering amplitudes of the coupled channel systems $D_{s}\pi$ and $\pi\pi$.
However, neither the tree-level nor both the tree-level and $\pi\pi$ coupled channel scattering contributions can describe the experimental data well, which indicates that the $D_{s}\pi$ coupled channel scattering is required.
We find that the $T_{c\bar{s}}$ and $f_{0}(500)$ resonances are dynamically generated from the pseudoscalar-pseudoscalar meson interaction within the chiral unitary approach.
In addition, we find the corresponding poles and also calculate the $S$-wave scattering length for the $DK$ channel, which is on the same scale as the result extracted from the experiment.
We propose that the $T_{c\bar{s}}$ and $f_{0}(500)$ resonances in the $D_{s1}(2460)^{+} \rightarrow D_{s}^{+}\pi^{+}\pi^{-}$ reaction are hadronic molecular type particles.

\end{abstract}
\maketitle

\section{Introduction}\label{sec:Introduction}

Recently, the LHCb Collaboration reported the measurement of $D_{s1}(2460)^{+} \rightarrow D_{s}^{+}\pi^{+}\pi^{-}$ reaction \cite{LHCb:2024iuo}.
The mass distributions of $D_{s}^{+}\pi^{+}$ and $\pi^{+}\pi^{-}$ can be described in the model by including the $f_{0}(500)$, $f_{0}(980)$, and $f_{2}(1270)$ resonances. However, the contributions from $f_{0}(980)$ and $f_{2}(1270)$ are unexpectedly large.
Another model, that includes the $f_{0}(500)$ resonance and a new open-charm tetraquark state $T_{c\bar{s}}$ with its isospin partner, can also describe the experimental data well.
The high significance is an indication that the existence of this new $T_{c\bar{s}}$ state is credible.
The measured mass and width are
\begin{equation*}
\begin{aligned}
M_{T_{c\bar{s}}}&=(2327\pm13\pm13)~\text{MeV}, \\
\Gamma_{T_{c\bar{s}}}&=(96\pm16^{+170}_{-23})~\text{MeV},
\end{aligned}
\end{equation*}
respectively. Its mass is slightly below the $DK$ threshold and it has a favoured spin parity with $0^{+}$.
We assume that it is correlated with the $DK$ component of the $S$-wave, which seems reasonable.
Besides, this decay provides the latest experimental data for a deeper understanding of the $f_{0}(500)$ resonance.
Therefore, we investigate the scalar resonance contribution in this reaction based on the final state interaction to explore the source of $T_{c\bar{s}}$ state and the properties of the $f_{0}(500)$ resonance in the present work.

Indeed, this is not the first instance that the open-charm tetraquark state is experimentally discovered.
It is well known that the LHCb Collaboration has found a tetraquark state $T_{c\bar{s}}(2900)$ in the $D_{s}^{+}\pi^{-}$ and $D_{s}^{+}\pi^{+}$ invariant mass distributions in the $B^{0}\rightarrow \bar{D}^{0}D_{s}^{+}\pi^{-}$ and $B^{+}\rightarrow D^{-}D_{s}^{+}\pi^{+}$ decays \cite{LHCb:2022sfr,LHCb:2022lzp}.
The new $T_{c\bar{s}}$ state discovered this time is very similar to it, both appearing in the $D_{s}^{+}\pi^{-}$ and $D_{s}^{+}\pi^{+}$ mass distributions, with only differences in mass and width.
Many theoretical works suggest that it is a $D^{*}K^{*}$ molecular candidate \cite{Molina:2010tx,Agaev:2022duz,Agaev:2022eyk,Duan:2023qsg,Lyu:2023ppb,Duan:2023lcj,Wang:2023hpp,Lyu:2023aqn,Lyu:2024wxa}, considering the mass of $T_{c\bar{s}}(2900)$ being close to the $D^{*}K^{*}$ threshold. Besides, in the quark model it can be explained as a compact tetraquark state in Refs. \cite{Liu:2022hbk,Dmitrasinovic:2023eei,Lian:2023cgs,Yang:2023evp,Ortega:2023azl,Vogt:2024fky,Wan:2024ykm}, such as the lowest $1S$-wave tetraquark state \cite{Liu:2022hbk}.
Furthermore, some theoretical studies suggest that these structures may be explained by the threshold effect, triangle singularity, or other dynamic mechanisms \cite{Ke:2022ocs,Molina:2022jcd,Molina:2023ghu,Luo:2023vha}, rather than a real particle.
In addition, the open charm states with strangeness in the three-meson interaction have been studied in Ref. \cite{Malabarba:2022pdo}.
These works enlighten us to think about how to understand this new $T_{c\bar{s}}$ state.

Going back to 2016, the D0 Collaboration discovered the $X(5568)$ state in the $B_{s}^{0}\pi^{\pm}$ mass distribution based on the $p\bar{p}$ collision data \cite{D0:2016mwd}.
Its mass is close to the thresholds of $B_{s}\pi$ and $B\bar{K}$, which may be a partner state of $T_{c\bar{s}}$ in the bottom sector.
In the year that the $X(5568)$ state was discovered, there were many different theoretical perspectives on this state \cite{Agaev:2016mjb,Wang:2016mee,Wang:2016tsi,Chen:2016mqt,Zanetti:2016wjn,Xiao:2016mho,Agaev:2016ijz,Liu:2016xly,Liu:2016ogz,Agaev:2016urs,Wang:2016wkj,Stancu:2016sfd,Burns:2016gvy,Tang:2016pcf,Guo:2016nhb,Lu:2016zhe,Albaladejo:2016eps,Chen:2016npt,Albuquerque:2016nlw,Kang:2016zmv,Lang:2016jpk,Chen:2016ypj,Lu:2016kxm,Agaev:2016ifn,Goerke:2016hxf,Yang:2016sws}, and the debate continues to this day. 
It is interesting that, for $D_{s}\pi$ and $DK$ coupled system, $D_{s}^{*}\rho$ and $D^{*}K^{*}$ coupled system, as well as $B_{s}\pi$ and $B\bar{K}$ coupled system, their diagonal interactions are OZI-suppressed, and only the non-diagonal elements are allowed to interact by exchanging a meson. The detailed analysis can be found in Refs. \cite{Molina:2010tx,Albaladejo:2016eps}.
When taking the $X(5568)$ being a physical state, there should be one molecular state formed by $D_{s}\pi$ and $DK$ channels with mass of $2210.8$ MeV and width of $50.1$ MeV, just as mentioned in Ref. \cite{Albaladejo:2016eps}.
Under this hypothesis, the theoretical predictions are not consistent with the experimental measurements \cite{LHCb:2024iuo}, and this proves the view of Ref. \cite{Albaladejo:2016eps} that $X(5568)$ is not a physical state.
Besides in 2009, the authors of Ref. \cite{Guo:2009ct} calculated the $S$-wave scattering lengths for the $D_{s}\pi$ and $DK$ channels, and also predicted possible bound states by using the chiral perturbation theory. The predictions show some differences from the LHCb Collaboration's measurement \cite{LHCb:2024iuo}.

In the field of particle physics, the exotic hadronic states, including molecular states, compact multiquark states, glueballs, hybrids, and other novel forms, are always hot topics in recent years \cite{Liu:2013waa,Hosaka:2016pey,Chen:2016qju,Richard:2016eis,Lebed:2016hpi,Olsen:2017bmm,Guo:2017jvc,Liu:2019zoy,Brambilla:2019esw,Meng:2022ozq,Chen:2022asf,Liu:2024uxn}.
In particular, the tetraquark composition of $T_{c\bar{s}}$ cannot be described by the traditional $q\bar{q}$ picture, which is a good candidate for the exotic state.
We start from the $D_{s1}(2460)^{+}$ decay at the quark level in order to provide the coupled channels that contribute to the final states $D_{s}^{+}\pi^{+}\pi^{-}$.
Then, taking into account the final state interaction, the decay amplitude is calculated by using the chiral unitary approach (ChUA).
We know that the $f_{0}(500)$ resonance can be dynamically generated in the $S$-wave interactions of $\pi\pi$ with its coupled channels \cite{Oller:1997ti}.
Furthermore, the authors of Ref. \cite{Tang:2023yls} studied the $D_{s1}(2460)^{+} \rightarrow D_{s}^{+}\pi^{+}\pi^{-}$ decay, by treating the $D_{s1}(2460)^{+}$ as a $D^{*}K$ hadronic molecule, and predicted two structures in the $\pi^{+}\pi^{-}$ invariant mass spectrum with the $f_{0}(500)$ contribution.
It is to be tested whether the calculated amplitude can describe the mass distributions of $D_{s}^{+}\pi^{+}$ and $\pi^{+}\pi^{-}$, and whether the $T_{c\bar{s}}$ state can be reproduced in the interactions of the $D_{s}\pi$ and $DK$ channels.

This work is organized as follows.
After the Introduction, we derive the amplitude of the $D_{s1}(2460)^{+} \rightarrow D_{s}^{+}\pi^{+}\pi^{-}$ reaction based on the final state interaction in Sec. \ref{sec:Formalism}. 
Then the numerical results are presented in Sec. \ref{sec:Results}.
At last, a brief summary is given in Sec. \ref{sec:Summary}.

\section{Formalism}\label{sec:Formalism}

In the present work, we investigate the $D_{s1}(2460)^{+} \rightarrow D_{s}^{+}\pi^{+}\pi^{-}$ reaction through the final state interaction. 
Since the $D_{s1}(2460)^{+}$ state has a spin $1$ and decays to three-particle final state $D_{s}^{+}$, $\pi^{+}$, and $\pi^{-}$, which have spins $0$, we need the polarization vector of $D_{s1}(2460)^{+}$ and the momentum of a final meson to form a scalar.
According to Refs. \cite{Liang:2016hmr,Li:2024uwu}, we can construct the amplitude structure of the $D_{s1}(2460)^{+} \rightarrow D_{s}^{+}\pi^{+}\pi^{-}$ decay as
\begin{equation}
\begin{aligned}
t=&(\vec{\epsilon}_{D_{s1}(2460)^{+}} \cdot \vec{p}_{D_{s}^{+}})\tilde{t}_{\pi^{+}\pi^{-}}e^{i\delta_{1}} \\
&+(\vec{\epsilon}_{D_{s1}(2460)^{+}} \cdot \vec{p}_{\pi^{+}})\tilde{t}_{D_{s}^{+}\pi^{-}}e^{i\delta_{2}} \\
&+(\vec{\epsilon}_{D_{s1}(2460)^{+}} \cdot \vec{p}_{\pi^{-}})\tilde{t}_{D_{s}^{+}\pi^{+}}.
\end{aligned}
\label{eq:t}
\end{equation}
In the first item, the $D_{s}^{+}$ meson has the orbital angular momentum $L=1$.
The symbol $\tilde{t}_{\pi^{+}\pi^{-}}$ represents the $S$-wave amplitude from the interactions between $\pi\pi$ and its coupled channels.
In principle, there will also be the $D_{s}^{+}$ interacting with pions, but the interaction is very weak and negligible since it will involve a $P$-wave \cite{Liang:2016hmr}.
In the second and third terms, the $\pi^{+}$ and $\pi^{-}$ mesons have orbital angular momenta $L=1$, and $\tilde{t}_{D_{s}^{+}\pi^{-}}$ and $\tilde{t}_{D_{s}^{+}\pi^{+}}$ represent the $S$-wave amplitudes from the interactions $D_{s}^{+}\pi^{-}$, $D_{s}^{+}\pi^{+}$ with their coupled channels, respectively.
Besides, we fix the phase of the third term and introduce two relative phases $\delta_{1}$ and $\delta_{2}$ into the first two terms, taking them as free parameters, as done in Ref. \cite{Xie:2016evi}.

For obtaining the proper combination of final state mesons of $D_{s1}(2460)^{+}$ decay, we introduce the quark level $q\bar{q}$ matrix, i.e.,
\begin{equation}
\begin{aligned}
M=\left(\begin{array}{llll}{u \bar{u}} & {u \bar{d}} & {u \bar{s}} & {u \bar{c}} \\ {d \bar{u}} & {d \bar{d}} & {d \bar{s}} & {d \bar{c}} \\ {s \bar{u}} & {s \bar{d}} & {s \bar{s}} & {s \bar{c}} \\ {c \bar{u}} & {c \bar{d}} & {c \bar{s}} & {c \bar{c}} \end{array}\right).
\end{aligned}
\label{eq:M}
\end{equation}
The properties of this matrix are given in Refs. \cite{Liang:2014tia,Liang:2016hmr}.
The quark components of the initial $D_{s1}(2460)^{+}$ state are $c\bar{s}$ which located in the fourth row and third column of the $M$ matrix.
So the meson combinations that occur in its three-body decay can be found in $(M\cdot M\cdot M)_{43}$, $(M\cdot M)_{43}\text{Tr}(M)$, $(M)_{43}\text{Tr}(M)\text{Tr}(M)$, $(M)_{43}\text{Tr}(M\cdot M)$, and so on.
Only the first term is the OZI-allowed contribution, while the other terms are OZI-suppressed because they contain the trace of one or two $M$ matrices.
As suggested by the experiment, this decay process gets the main contribution from the $f_{0}(500)$ and $T_{c\bar{s}}$ states.
So we only consider the interactions between $\pi\pi$ and $K\bar{K}$ channels, as well as $D_{s}\pi$ and $DK$ channels.
In this case, we find that the contributions come from the items $(M\cdot M\cdot M)_{43}$ and $(M)_{43}\text{Tr}(M\cdot M)$, with no contributions from other items.
Then we write the matrix $M$ in terms of the pseudoscalar mesons
\begin{equation}
\begin{aligned}
P=\left(\begin{array}{cccc}
\frac{\eta}{\sqrt{3}}+\frac{\eta^{\prime}}{\sqrt{6}}+\frac{\pi^0}{\sqrt{2}} & \pi^{+} & K^{+} & \bar{D}^0  \\
\pi^{-} & \frac{\eta}{\sqrt{3}}+\frac{\eta^{\prime}}{\sqrt{6}}-\frac{\pi^0}{\sqrt{2}} & K^0 & D^{-}  \\
K^{-} & \bar{K}^0 & -\frac{\eta}{\sqrt{3}}+\sqrt{\frac{2}{3}}\eta^{\prime} & D_s^{-} \\
D^0 & D^{+} & D_s^{+} & \eta_c 
\end{array}\right).
\end{aligned}
\label{eq:P}
\end{equation}

Thus we can replace the $M$ matrix by the $P$ matrix and obtain the contributions of the $(M\cdot M\cdot M)_{43}$, which include the following combination of three mesons
\begin{equation}
\begin{aligned}
|H^{(1)}\rangle
=D_{s}^{+}K^{+}K^{-}+D_{s}^{+}K^{0}\bar{K}^{0}+\pi^{+}D^{0}K^{0}+\pi^{-}D^{+}K^{+}.
\end{aligned}
\label{eq:H1}
\end{equation}
Note that we only keep terms contributing to the final states $D_{s}^{+}\pi^{+}\pi^{-}$. 
For the term of $(M)_{43}\text{Tr}(M\cdot M)$, we have the following combination of three mesons
\begin{equation}
\begin{aligned}
|H^{(2)}\rangle
=2D_{s}^{+}\pi^{+}\pi^{-}+D_{s}^{+}\pi^{0}\pi^{0}+2D_{s}^{+}K^{+}K^{-}+2D_{s}^{+}K^{0}\bar{K}^{0}.
\end{aligned}
\label{eq:H2}
\end{equation}
\begin{figure}[htbp]
\begin{minipage}{0.75\linewidth}
\centering
\includegraphics[width=1\linewidth,trim=150 550 220 130,clip]{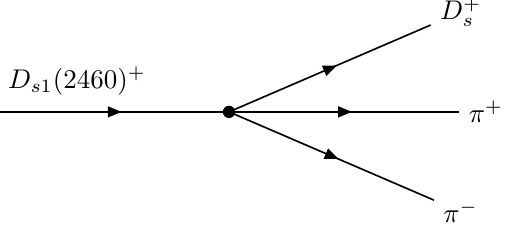} 
\label{fig:Tree1}
\end{minipage}
\begin{minipage}{0.8\linewidth} 
\centering 
\includegraphics[width=1\linewidth,trim=150 540 190 130,clip]{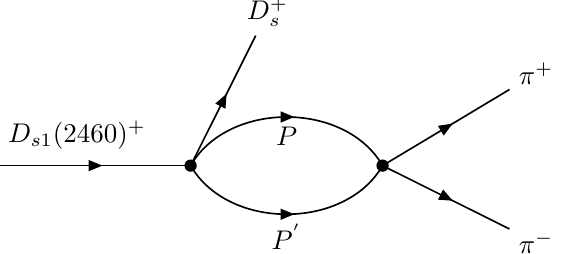}
\label{fig:Scattering1}  
\end{minipage}
\caption{Production of $D_{s}^{+}\pi^{+}\pi^{-}$ in the $D_{s1}(2460)^{+}$ decay through the tree-level (above) and rescattering of the $PP^{'}$ pair (below), where $PP^{'}$ represents the $\pi^{+}\pi^{-}$, $\pi^{0}\pi^{0}$, $K^{+}K^{-}$, and $K^{0}\bar{K}^{0}$ channels.}
\label{fig:Ds2460decay1}
\end{figure}

\begin{figure}[htbp]
\begin{minipage}{0.75\linewidth}
\centering
\includegraphics[width=1\linewidth,trim=150 550 220 130,clip]{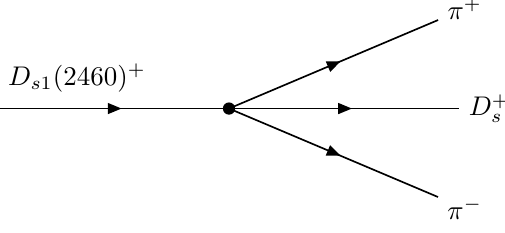} 
\label{fig:Tree2}
\end{minipage}
\begin{minipage}{0.8\linewidth} 
\centering 
\includegraphics[width=1\linewidth,trim=150 540 190 130,clip]{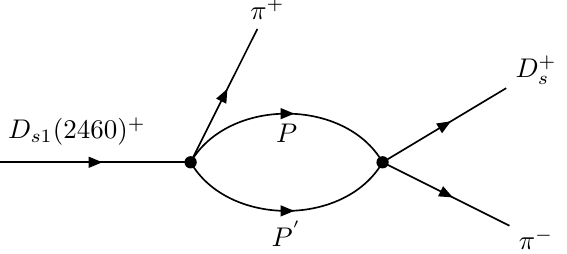}
\label{fig:Scattering2}  
\end{minipage}
\caption{Production of $\pi^{+}D_{s}^{+}\pi^{-}$ in the $D_{s1}(2460)^{+}$ decay through the tree-level (above) and rescattering of the $PP^{'}$ pair (below), where $PP^{'}$ represents the $D_{s}^{+}\pi^{-}$ and $D^{0}{K}^{0}$ channels.}
\label{fig:Ds2460decay2}
\end{figure}

\begin{figure}[htbp]
\begin{minipage}{0.75\linewidth}
\centering
\includegraphics[width=1\linewidth,trim=150 550 220 130,clip]{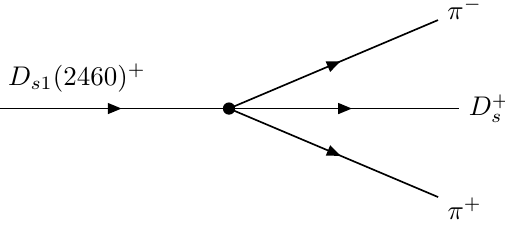} 
\label{fig:Tree3}
\end{minipage}
\begin{minipage}{0.8\linewidth} 
\centering 
\includegraphics[width=1\linewidth,trim=150 540 190 130,clip]{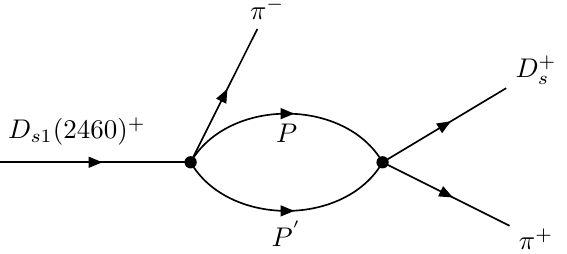}
\label{fig:Scattering3}  
\end{minipage}
\caption{Production of $\pi^{-}D_{s}^{+}\pi^{+}$ in the $D_{s1}(2460)^{+}$ decay through the tree-level (above) and rescattering of the $PP^{'}$ pair (below), where $PP^{'}$ represents the $D_{s}^{+}\pi^{+}$ and $D^{+}{K}^{+}$ channels.}
\label{fig:Ds2460decay3}
\end{figure}

So if the $D_{s}^{+}$ meson is in the $P$-wave, we will have $\pi^{+}\pi^{-}$, $\pi^{0}\pi^{0}$, $K^{+}K^{-}$, and $K^{0}\bar{K}^{0}$ channels in the primary step which will undergo final state interaction to produce a $\pi^{+}\pi^{-}$ pair.
Fig. \ref{fig:Ds2460decay1} shows the tree-level and the rescattering contributions in this process.
The corresponding amplitude of the $S$-wave is given by
\begin{equation}
\begin{aligned}
\tilde{t}_{\pi^{+}\pi^{-}}(M_{23})
=&V_{1}\left[G_{K^{+}K^{-}}(M_{23})T_{K^{+}K^{-} \rightarrow \pi^{+}\pi^{-}}(M_{23}) \right.\\ & \left.
+G_{K^{0}\bar{K}^{0}}(M_{23})T_{K^{0}\bar{K}^{0} \rightarrow \pi^{+}\pi^{-}}(M_{23})\right] \\
+&V_{2}\left[2+2G_{\pi^{+}\pi^{-}}(M_{23})T_{\pi^{+}\pi^{-} \rightarrow \pi^{+}\pi^{-}}(M_{23}) \right.\\ & \left.
+\sqrt{2}G_{\pi^{0}\pi^{0}}(M_{23})T_{\pi^{0}\pi^{0} \rightarrow \pi^{+}\pi^{-}}(M_{23}) \right.\\ & \left.
+2G_{K^{+}K^{-}}(M_{23})T_{K^{+}K^{-} \rightarrow \pi^{+}\pi^{-}}(M_{23}) \right.\\ & \left.
+2G_{K^{0}\bar{K}^{0}}(M_{23})T_{K^{0}\bar{K}^{0} \rightarrow \pi^{+}\pi^{-}}(M_{23})\right],
\end{aligned}
\label{eq:t1}
\end{equation}
where $V_{1}$ and $V_{2}$ are the decay strengths from Eqs. \eqref{eq:H1} and \eqref{eq:H2}, respectively. 
We take them as constants and determine them by fitting the experimental data.
The symbol $M_{ij}$ is the energy of two particles in the center-of-mass frame, where the lower indices $(i,j)=(1,2,3)$ denote the three final states of $D_{s}^{+}$, $\pi^{+}$, and $\pi^{-}$, respectively.
Note that there is a factor of $2$ in the term related to the identical particles $\pi^{0}\pi^{0}$ fields, which has been cancelled by the factor of $1/2$ in their loop function.
The factor $\sqrt{2}$ in the $\pi^{0}\pi^{0}$ channel has the same root, as in the chiral unitary approach the amplitudes are evaluated with the unitary normalization $|\pi^{0}\pi^{0}\rangle\rightarrow\frac{1}{\sqrt{2}}|\pi^{0}\pi^{0}\rangle$.\footnote{We are very grateful to the referee for the valuable comments and very insightful remarks.}
For more discussion see Refs. \cite{Oller:1997ti,Dias:2016gou}.
The $G$ is the loop function of two mesons, and the $T$ is the scattering amplitude, which will be introduced later.
In the same way, as depicted in Fig. \ref{fig:Ds2460decay2}, if the final state $\pi^{+}$ carries the $P$-wave, the amplitude of the $S$-wave is given by
\begin{equation}
\begin{aligned}
\tilde{t}_{D_{s}^{+}\pi^{-}}(M_{13})
=&V_{1}G_{D^{0}K^{0}}(M_{13})T_{D^{0}K^{0} \rightarrow D_{s}^{+}\pi^{-}}(M_{13}) \\
+&V_{2}\left[2+2G_{D_{s}^{+}\pi^{-}}(M_{13})T_{D_{s}^{+}\pi^{-} \rightarrow D_{s}^{+}\pi^{-}}(M_{13}) \right].
\end{aligned}
\label{eq:t2}
\end{equation}
In the third term of Eq. \eqref{eq:t}, the final state $\pi^{-}$ carries the $P$-wave.
After the final state interaction in Fig. \ref{fig:Ds2460decay3}, the amplitude of the $S$-wave is given by
\begin{equation}
\begin{aligned}
\tilde{t}_{D_{s}^{+}\pi^{+}}(M_{12})
=&V_{1}G_{D^{+}K^{+}}(M_{12})T_{D^{+}K^{+} \rightarrow D_{s}^{+}\pi^{+}}(M_{12}) \\
+&V_{2}\left[2+2G_{D_{s}^{+}\pi^{+}}(M_{12})T_{D_{s}^{+}\pi^{+} \rightarrow D_{s}^{+}\pi^{+}}(M_{12}) \right].
\end{aligned}
\label{eq:t3}
\end{equation}
At present there are three variables $M_{12}$, $M_{13}$, and $M_{23}$ in the amplitude of Eq. \eqref{eq:t}, which are not completely independent.
They satisfy the following constraint condition, which means that only two of them are independent
\begin{equation}
\begin{aligned}
M_{12}^{2}+M_{13}^{2}+M_{23}^{2}=m_{D_{s1}(2460)^{+}}^{2}+m_{D_{s}^{+}}^{2}+m_{\pi^{+}}^{2}+m_{\pi^{-}}^{2}.
\end{aligned}
\label{eq:M12M13M23}
\end{equation}

In this paper we calculate the rescattering amplitudes in the isospin basis.
The $S$-wave interactions of the two channels $\pi\pi$ and $K\bar{K}$ in the isospin $I=0$ system, and of the two channels $D_{s}\pi$ and $DK$ in the $I=1$ system are considered.
Then we decompose the amplitudes of the physical states into those of the isospin states
\begin{equation}
\begin{aligned}
& T_{\pi^{+}\pi^{-}\rightarrow \pi^{+}\pi^{-}}=\frac{2}{3}T_{\pi\pi\rightarrow \pi\pi}^{I=0}, \quad
T_{\pi^{0}\pi^{0}\rightarrow \pi^{+}\pi^{-}}=\frac{\sqrt{2}}{3}T_{\pi\pi\rightarrow \pi\pi}^{I=0}, \\
& T_{K^{+}K^{-}\rightarrow \pi^{+}\pi^{-}}=\frac{1}{\sqrt{3}}T_{K\bar{K}\rightarrow \pi\pi}^{I=0}, \quad
T_{K^{0}\bar{K}^{0}\rightarrow \pi^{+}\pi^{-}}=\frac{1}{\sqrt{3}}T_{K\bar{K}\rightarrow \pi\pi}^{I=0}, \\
& T_{D^{0}K^{0}\rightarrow D_{s}^{+}\pi^{-}}=-T_{DK\rightarrow D_{s}\pi}^{I=1}, \quad
T_{D^{+}K^{+}\rightarrow D_{s}^{+}\pi^{+}}=-T_{DK\rightarrow D_{s}\pi}^{I=1},
\end{aligned}
\label{eq:isospinstates}
\end{equation}
where we have used the phase convention $\left|\pi^{+}\rangle=-\right|1,1\rangle$, $\left|K^{-}\rangle=-\right| 1/2,-1/2\rangle$, and $\left|D^{0}\rangle=-\right| 1/2,-1/2\rangle$ \cite{Wang:2022xqc,Wang:2022aga}.
Here we omit the contribution of the $\pi\pi$ channel with isospin $I=2$.

Then the two-body scattering amplitudes of the coupled channels can be obtained by solving the Bethe-Salpeter equation factorized on shell \cite{Oller:1997ti,Oset:1997it,Oller:1997ng}
\begin{equation}
\begin{aligned}
T = [1-VG]^{-1}V.
\end{aligned}
\label{eq:BSE}
\end{equation}
Furthermore, the $G$ in Eqs.~(\ref{eq:t1}-\ref{eq:t3}) and Eq.~\eqref{eq:BSE} is a diagonal matrix formed by the loop function of two intermediate particles.
The expression for the three-momentum cutoff of its elements can be written as \cite{Oller:1997ti}
\begin{equation}
\begin{aligned}
G _ { ii } ( s ) = \int _ { 0 } ^ { q _ { \max } } \frac { q ^ { 2 } d q } { ( 2 \pi ) ^ { 2 } } \frac { \omega _ { 1 } + \omega _ { 2 } } { \omega _ { 1 } \omega _ { 2 } \left[ s - \left( \omega _ { 1 } + \omega _ { 2 } \right) ^ { 2 } + i \epsilon \right] },
\end{aligned}
\label{eq:GCO}
\end{equation}
where $\omega_{i}(\vec{q})=\sqrt{\vec{q}\,^2+m_{i}^2}$, $m_{1}$ and $m_{2}$ are the masses of the two particles in the coupled system, and $q_{max}$ is a free parameter.
Another expression for the method of dimensional regularization is given by \cite{Oller:1998zr,Oller:2000fj,Gamermann:2006nm,Alvarez-Ruso:2010rqm,Guo:2016zep}
\begin{equation}
\begin{aligned}
G_{ii}(s)=& \frac{1}{16 \pi^{2}}\left\{a_{i}(\mu)+\ln \frac{m_{1}^{2}}{\mu^{2}}+\frac{m_{2}^{2}-m_{1}^{2}+s}{2 s} \ln \frac{m_{2}^{2}}{m_{1}^{2}}\right.\\
&+\frac{q_{cmi}(s)}{\sqrt{s}}\left[\ln \left(s-\left(m_{2}^{2}-m_{1}^{2}\right)+2 q_{cmi}(s) \sqrt{s}\right)\right.\\
&+\ln \left(s+\left(m_{2}^{2}-m_{1}^{2}\right)+2 q_{cmi}(s) \sqrt{s}\right) \\
&-\ln \left(-s-\left(m_{2}^{2}-m_{1}^{2}\right)+2 q_{cmi}(s) \sqrt{s}\right) \\
&\left.\left.-\ln \left(-s+\left(m_{2}^{2}-m_{1}^{2}\right)+2 q_{cmi}(s) \sqrt{s}\right)\right]\right\},
\end{aligned}
\label{eq:GDR}
\end{equation}
where the regularization scale $\mu$ is a free parameter, and the subtraction constant $a_{i}(\mu)$ is dependent on the chosen of the regularization scale $\mu$. 
Besides, $q_{cmi}(s)$ is three-momentum of the particle in the center-of-mass frame
\begin{equation}
\begin{aligned}
q_{cmi}(s)=\frac{\lambda^{1 / 2}\left(s, m_{1}^{2}, m_{2}^{2}\right)}{2 \sqrt{s}},
\end{aligned}
\label{eq:qcmi}
\end{equation}
with the K\"all\'en triangle function $\lambda(a, b, c)=a^{2}+b^{2}+c^{2}-2(a b+a c+b c)$.
Since the three-momentum cutoff method still exhibits a singularity above the threshold, we adopt the dimensional regularization method in this paper.
For the subtraction constant $a_{i}(\mu)$, we match the values of $G$ from these two methods at the threshold to determine it in a given channel, as done in Refs. \cite{Oset:2001cn,Montana:2022inz,Liang:2023ekj,Wang:2024yzb}
\begin{equation}
\begin{aligned}
a_{i}(\mu)=16\pi^{2}[G^{CO}(s_{thr},q_{max})-G^{DR}(s_{thr},\mu)],
\end{aligned}
 \label{eq:ai}
\end{equation}
where we take $q_{max}=\mu$, and $G^{CO}$ and $G^{DR}$ are given by Eqs. \eqref{eq:GCO} and \eqref{eq:GDR}, respectively.
The value of the regularization scale $\mu$ will be discussed in detail in Sec. \ref{sec:Results}.

In addition, the matrix $V$ in Eq. \eqref{eq:BSE} denotes the $S$-wave interaction potentials for the coupled channels of $\pi\pi$ with isospin $I=0$ and the coupled channels of $D_{s}\pi$ with $I=1$, whose elements are derived from the chiral Lagrangian.
For the $\pi\pi$ and its coupled channel $K\bar{K}$, we have \cite{Ahmed:2020kmp}
\begin{equation}
\begin{aligned}
&V_{\pi\pi\rightarrow\pi\pi}^{I=0}=-\frac{1}{f^2}\left(s-\frac{1}{2}m_{\pi}^{2}\right), \\
&V_{\pi\pi\rightarrow K\bar{K}}^{I=0}=-\frac{3s}{2\sqrt{12}f^2}, \\
&V_{K\bar{K}\rightarrow K\bar{K}}^{I=0}=-\frac{3s}{4f^2},
\end{aligned}
\label{eq:V1}
\end{equation}
where we take $f =0.093$ GeV, which is the pion decay constant.
For the $D_{s}\pi$ and its coupled channel $DK$, we have \cite{Albaladejo:2016eps}
\begin{equation}
\begin{aligned}
&V_{D_{s}\pi\rightarrow D_{s}\pi}^{I=1}=0, \\
&V_{D_{s}\pi\rightarrow DK}^{I=1}=-\frac{1}{8f^2}\left(3s-\left(M_{1}^{2}+M_{2}^{2}+m_{1}^{2}+m_{2}^{2}\right)-\frac{\Delta_{1}\Delta_{2}}{s}\right), \\
&V_{DK\rightarrow DK}^{I=1}=0,
\end{aligned}
\label{eq:V2}
\end{equation}
where $\Delta_{i}=M_{i}^{2}-m_{i}^{2}$, $M_{i}$ and $m_{i}$ are the masses of the heavy and light mesons in the $i$-th channel, respectively.

Finally, the double differential width formula for a three-body decay is given by \cite{ParticleDataGroup:2024cfk}
\begin{equation}
\begin{aligned}
\frac{d^{2} \Gamma}{d M_{12}d M_{23}}=\frac{1}{(2 \pi)^{3}} \frac{1}{8 m_{D_{s1}(2460)^{+}}^{3}}M_{12} M_{23} \overline{\sum}\sum\left|t\right|^{2},
\end{aligned}
\label{eq:dGamma}
\end{equation}
and the $d \Gamma/d M_{12}$ and $d \Gamma/d M_{23}$ can be obtained by integrating over each of the invariant mass variables.

\section{Numerical results}\label{sec:Results}

In our theoretical model, there are five free parameters, i.e., the regularization scale $\mu$ in the loop function, the strengths of two decay mechanisms $V_{1}$ and $V_{2}$, and the two phases $\delta_{1}$ and $\delta_{2}$, that need to be determined. By a combined fitting of the mass distributions of $\pi^{+}\pi^{-}$, $D_{s}^{+}\pi^{+}$, and $D_{s}^{+}\pi^{+}$ with $M_{\pi^{+}\pi^{-}}>0.39$ GeV measured by LHCb Collaboration \cite{LHCb:2024iuo}, these parameters can be fixed. The fitted values of these free parameters are presented in Table \ref{tab:Parameters1}, with $\chi^{2}/dof.=98.37/(81-5)=1.29$.

\begin{table*}[htbp]
\centering
\renewcommand\tabcolsep{2.0mm}
\renewcommand{\arraystretch}{1.50}
\caption{Values of the parameters from the fit in the $D_{s1}(2460)^{+}$ decay.}
\begin{tabular*}{178mm}{@{\extracolsep{\fill}}c|ccccc}
\toprule[1.00pt]
\toprule[1.00pt]
Parameters&$\mu$&$V_{1}$&$V_{2}$&$\delta_{1}$&$\delta_{2}$ \\
\hline
Values&$1.078\pm0.104$ GeV&$50020.35\pm13150.27$&$-13300.65\pm1949.59$&$-0.41\pm0.02$ rad&$-0.06\pm0.03$ rad \\
\bottomrule[1.00pt]
\bottomrule[1.00pt]
\end{tabular*}
\label{tab:Parameters1}
\end{table*}

\begin{figure*}[htbp]
\begin{minipage}{0.33\linewidth}
\centering
\includegraphics[width=1\linewidth,trim=0 0 0 0,clip]{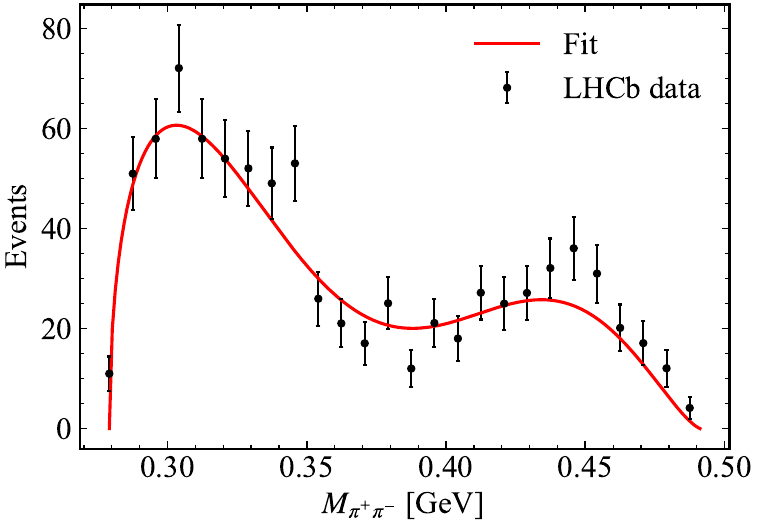} 
\label{fig:Figs1}
\end{minipage}
\begin{minipage}{0.33\linewidth}
\centering
\includegraphics[width=1\linewidth,trim=0 0 0 0,clip]{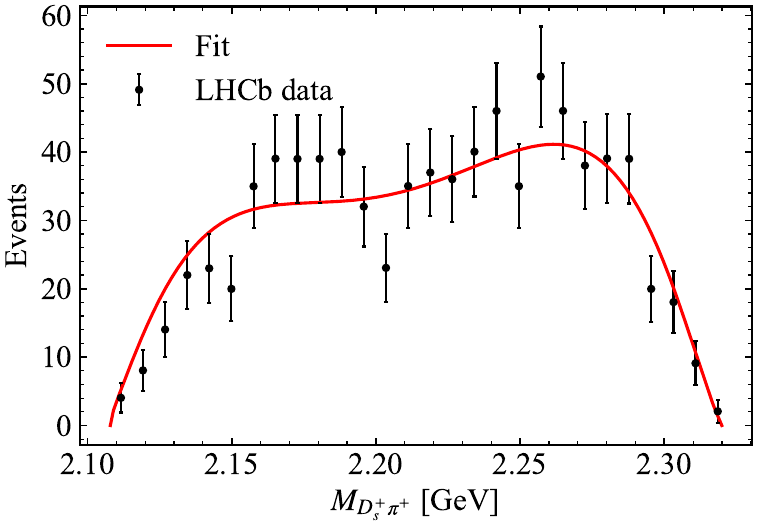} 
\label{fig:Figs2}
\end{minipage}
\begin{minipage}{0.33\linewidth}
\centering
\includegraphics[width=1\linewidth,trim=0 0 0 0,clip]{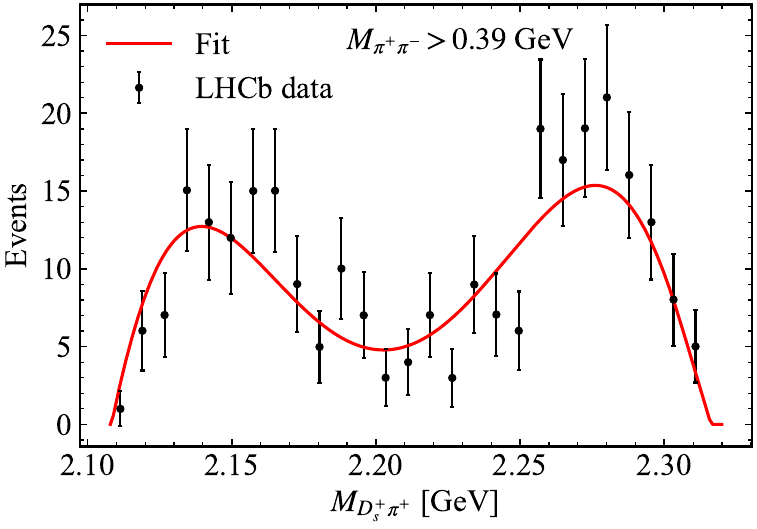} 
\label{fig:Figs3}
\end{minipage}
\begin{minipage}{0.33\linewidth}
\centering
\includegraphics[width=1\linewidth,trim=0 0 0 0,clip]{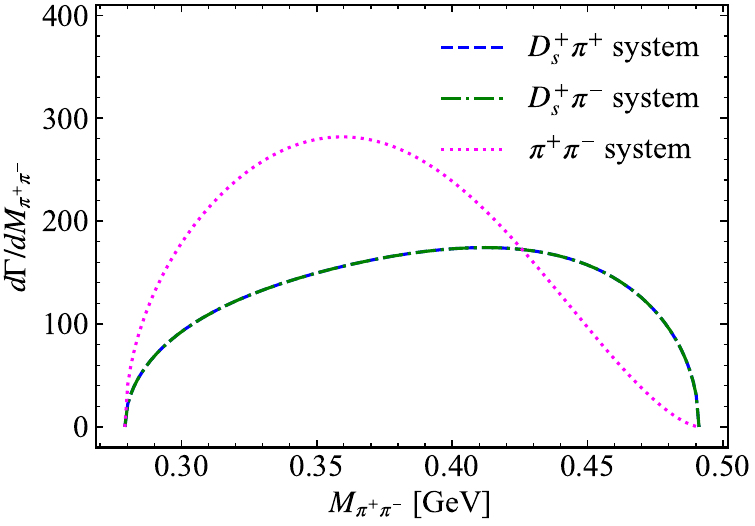} 
\label{fig:Figs21}
\end{minipage}
\begin{minipage}{0.33\linewidth}
\centering
\includegraphics[width=1\linewidth,trim=0 0 0 0,clip]{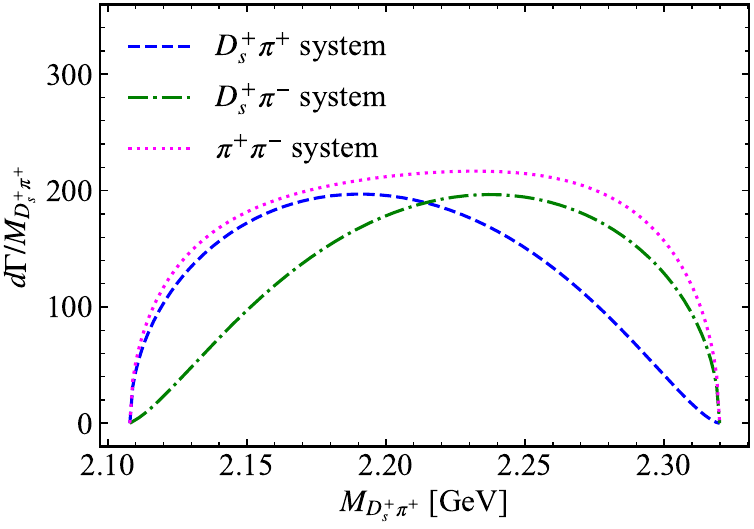} 
\label{fig:Figs22}
\end{minipage}
\begin{minipage}{0.33\linewidth}
\centering
\includegraphics[width=1\linewidth,trim=0 0 0 0,clip]{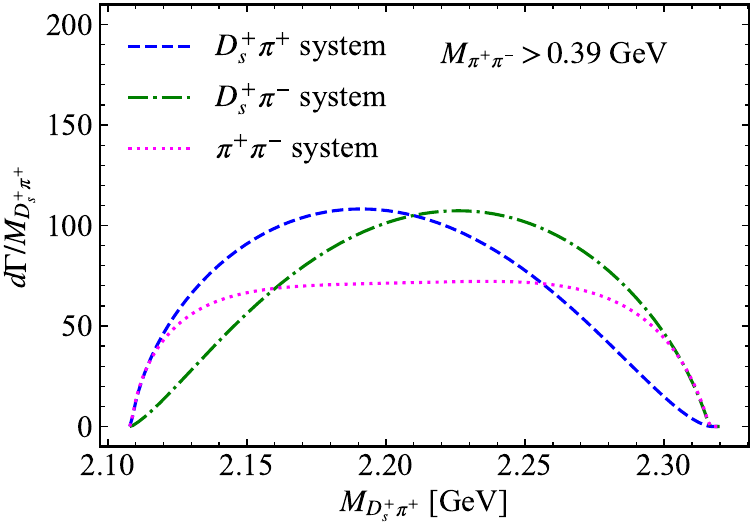} 
\label{fig:Figs23}
\end{minipage}
\caption{Invariant mass distributions of $\pi^{+}\pi^{-}$ and $D_{s}^{+}\pi^{+}$ in the $D_{s1}(2460)^{+} \rightarrow D_{s}^{+}\pi^{+}\pi^{-}$ decay (above row), and the corresponding $S$-wave components contribution (below row). The dot (black) points are the experimental data measured by the LHCb Collaboration, taken from Ref. \cite{LHCb:2024iuo}.}
\label{fig:Figs123}
\end{figure*}

\begin{table*}[htbp]
\centering
\renewcommand\tabcolsep{2.0mm}
\renewcommand{\arraystretch}{1.50}
\caption{The pole positions with different values of the regularisation scale $\mu$ (units: GeV).}
\begin{tabular*}{178mm}{@{\extracolsep{\fill}}c|ccccccc}
\toprule[1.00pt]
\toprule[1.00pt]
$\mu$&$1.0$&$1.1$&$1.2$&$1.3$&$1.4$&$1.5$&$1.6$ \\
\hline
$\sqrt{s_{p}}$&$2.409-0.055i$&$2.390-0.057i$&$2.373-0.059i$&$2.357-0.058i$&$2.342-0.057i$&$2.328-0.055i$&$2.316-0.053i$ \\
\bottomrule[1.00pt]
\bottomrule[1.00pt]
\end{tabular*}
\label{tab:Poles}
\end{table*}

\begin{figure*}[htbp]
\begin{minipage}{0.33\linewidth}
\centering
\includegraphics[width=1\linewidth,trim=0 0 0 0,clip]{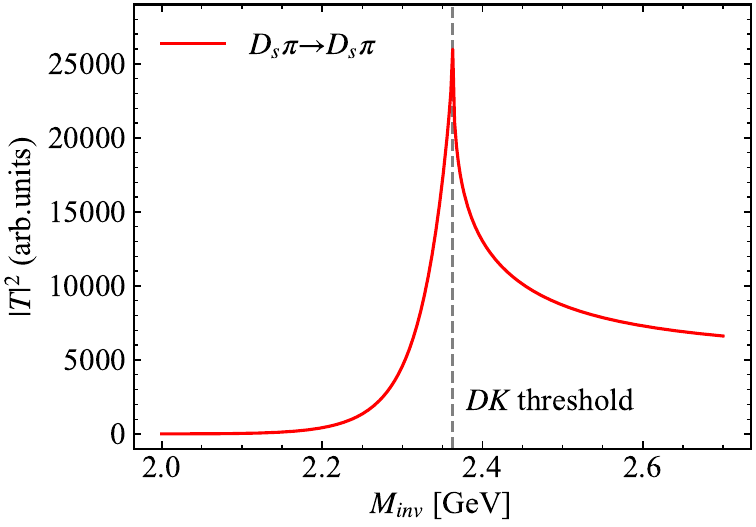} 
\label{fig:T11}
\end{minipage}
\begin{minipage}{0.33\linewidth}
\centering
\includegraphics[width=1\linewidth,trim=0 0 0 0,clip]{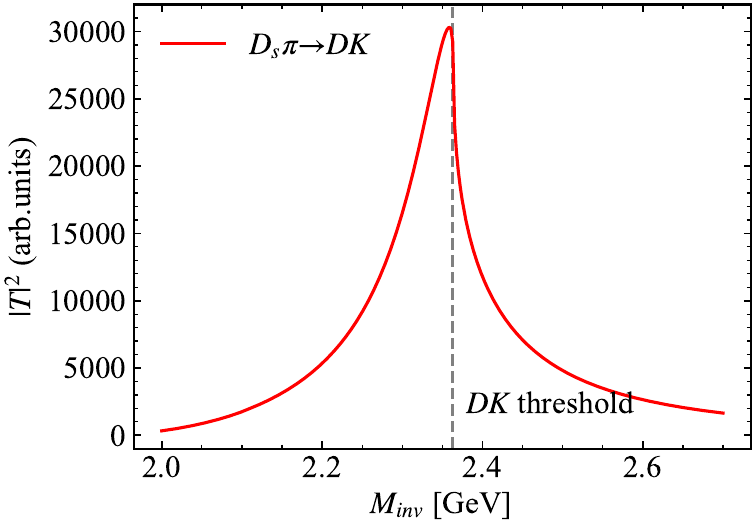} 
\label{fig:T12}
\end{minipage}
\begin{minipage}{0.33\linewidth}
\centering
\includegraphics[width=1\linewidth,trim=0 0 0 0,clip]{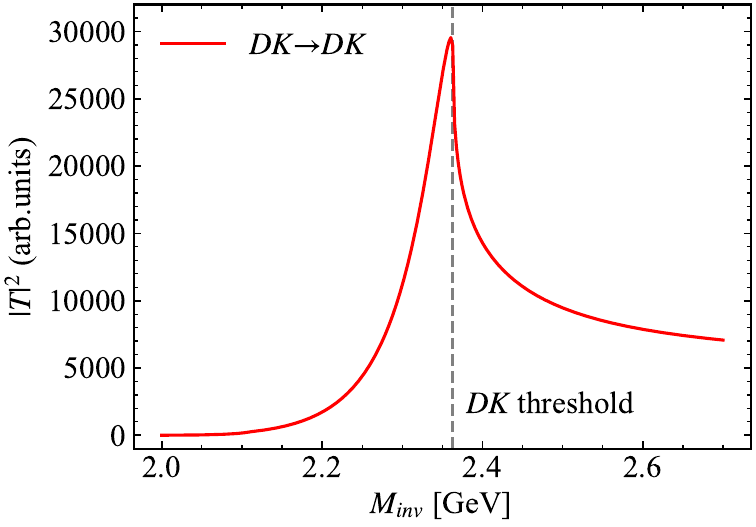} 
\label{fig:T22}
\end{minipage}
\begin{minipage}{0.33\linewidth}
\centering
\includegraphics[width=1\linewidth,trim=0 0 0 0,clip]{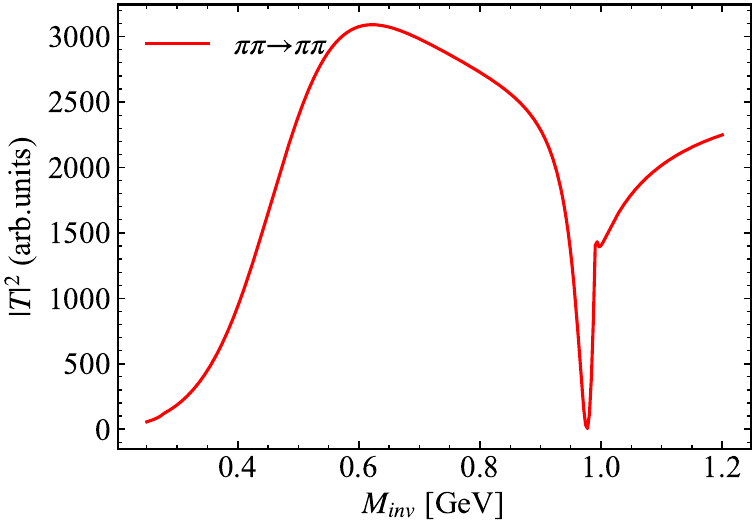} 
\label{fig:T1}
\end{minipage}
\begin{minipage}{0.33\linewidth}
\centering
\includegraphics[width=1\linewidth,trim=0 0 0 0,clip]{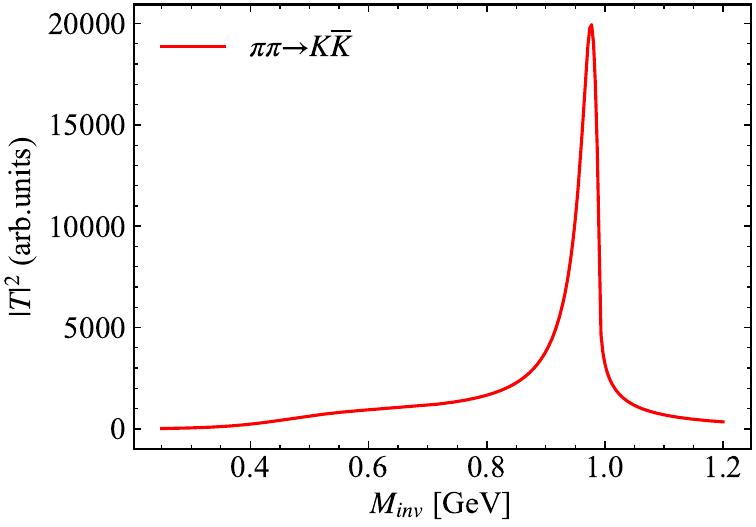} 
\label{fig:T2}
\end{minipage}
\begin{minipage}{0.33\linewidth}
\centering
\includegraphics[width=1\linewidth,trim=0 0 0 0,clip]{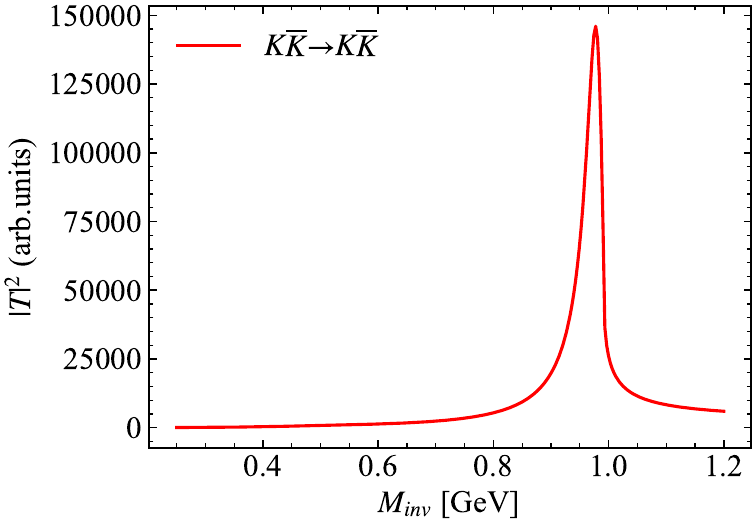} 
\label{fig:T3}
\end{minipage}
\caption{Modulus square of the $S$-wave amplitudes in the $D_{s}\pi$ (above row) and $\pi\pi$ (below row) coupled channels.}
\label{fig:TXX}
\end{figure*}

The fitted regularization scale $\mu$ is about $1$ GeV, which is very close to the empirical value for the $f_{0}(500)$ resonance.
It was found in Refs. \cite{Oller:1997ti,Xiao:2019lrj,Ahmed:2020kmp} that $q_{max}=0.931$ and $1.1$ GeV can simultaneously reproduce the $f_{0}(500)$, $f_{0}(980)$, and $a_{0}(980)$ resonances.
Note that in Eq. \eqref{eq:ai} we take $\mu=q_{max}$ to determine the value of the subtraction constant $a_{i}(\mu)$.
Using the fitted value of $\mu=1.078$ GeV, we obtain the subtraction constants from Eq. \eqref{eq:ai} as follows 
\begin{equation}
\begin{aligned}
&a_{\pi\pi}=-1.39, \quad a_{K\bar{K}}=-1.48, \\
&a_{D_{s}\pi}=-2.19, \quad a_{DK}=-2.05.
\end{aligned}
\label{eq:amu}
\end{equation}
For $V_{1}$ and $V_{2}$, they represent the strengths of the $D_{s1}(2460)^{+}$ decay contributed by the OZI-allowed and OZI-suppressed, respectively, so it is reasonable for the absolute of $V_{1}$ to be greater than the absolute of $V_{2}$.
In addition, they also include a global factor to match the experimental data, which means that the values we get have no other meaning.

The fitted results of the $\pi^{+}\pi^{-}$ and $D_{s}^{+}\pi^{+}$ invariant mass distributions are shown in the upper panel of Fig. \ref{fig:Figs123}.
Overall, our theoretical results are in good agreement with the experimental data.
The structures observed by the LHCb Collaboration can be reproduced.
There are two peaks in the invariant mass distribution of $\pi^{+}\pi^{-}$, where the one near the threshold is well described, while the later one is slightly smaller in our results.
From the components in Fig. \ref{fig:Figs123}, we can see that the first peak in the $\pi^{+}\pi^{-}$ mass distribution is mainly contributed by the $S$-wave amplitude obtained from $\pi\pi$ scattering, while the second peak is mainly contributed by the reflection of the $S$-wave amplitude from $D_{s}\pi$ scattering.
The two peaks in the $D_{s}^{+}\pi^{+}$ mass distribution are not obvious and come from the scattering amplitudes of the $D_{s}^{+}\pi^{+}$ and $D_{s}^{+}\pi^{-}$ systems, respectively, which are almost hidden in the $\pi^{+}\pi^{-}$ system reflection.
Therefore, by reducing the reflection of the $\pi^{+}\pi^{-}$ scattering amplitude, i.e. taking $M_{\pi^{+}\pi^{-}}>0.39$ GeV, the two peaks become clearly visible, and our results of the $D_{s}^{+}\pi^{+}$ mass distribution with $M_{\pi^{+}\pi^{-}}>0.39$ GeV are in good agreement with the experiment.
From the bottom panel of Fig. \ref{fig:Figs123}, we can see that the source of each peak structure is clear, whereas this feature is not present in the experimental fit.
In fact, to verify the necessity of $D_{s}\pi$ coupled channel scattering, we fit the cases with only tree-level, or both tree-level and $\pi\pi$ coupled channel scattering contributions. However the results are unable to reproduce the experimental data.
It should be noted that in our theoretical model, only the $S$-wave amplitude of the final state interaction is present, without the addition of any Breit-Wigner particles.

Next, we plot the modulus square of the amplitudes in the $D_{s}\pi$ and $\pi\pi$ coupled channel system at $\mu=1.078$ GeV, as shown in Fig. \ref{fig:TXX}.
For the $D_{s}\pi$ system, there are obvious bumps near the $DK$ threshold in all three amplitudes, which have cusp effects.
It is interesting that even if the interactions in the diagonal elements are zero, the peak structures can still be generated in these three amplitudes.
Then we find a pole at $\sqrt{s_{p}}=(2.394-0.057i)$ GeV on the second Riemann sheet.
The mass of this state is large than the $DK$ threshold of $2.363$ GeV, and also larger than the measured value of $2.327$ GeV, while the width $0.114$ GeV is agreement with the experimental result $(0.096\pm0.016^{+0.170}_{-0.023})$ GeV \cite{LHCb:2024iuo} within the uncertainties.
Then we calculate the coupling of the pole to each coupled channel using the formula \cite{Oller:2004xm,Guo:2006fu}
\begin{equation}
\begin{aligned}
T_{i j}(s)=\frac{g_{i} g_{j}}{s-s_{p}},
\end{aligned}
\label{eq:gij}
\end{equation}
where $g_i$ and $g_j$ are the couplings of the $i$-th and $j$-th channels, and $s_{p}$ is the square of the energy corresponding to the pole on the complex energy plane.
The couplings to $D_{s}\pi$ and $DK$ channels are determined as $g_{D_{s}\pi}=8.02$ GeV and $g_{DK}=9.37$ GeV, respectively.
This indicates that the $D_{s}\pi$ channel is also important.
Besides, we also find a pole $\sqrt{s_{p}}=(2.246-0.093i)$ GeV on the third Riemann sheet, which is close to the results of Ref. \cite{Guo:2009ct}.
In Table \ref{tab:Poles}, we present the variations of the pole with respect to the regularisation scale $\mu$.
Its real part decreases as $\mu$ increases, while the imaginary part remains almost unchanged.
If the value of $\mu$ is taken as $1.5$ GeV, the mass and width of this state are in good agreement with the experimental measurements.
In this case, if we set $\mu=0.931$ GeV for the $\pi\pi$ coupled channel system \cite{Xiao:2019lrj,Ahmed:2020kmp} and $\mu=1.5$ GeV for the $D_{s}\pi$ system, the fitted results can describe the $D_{s}^{+}\pi^{+}$ and $D_{s}^{+}\pi^{+}$ with $M_{\pi^{+}\pi^{-}}>0.39$ GeV invariant mass distributions, but not the $\pi^{+}\pi^{-}$ mass distribution.

In addition, we also calculate the scattering length of the $DK$ channel by using the following formula
\begin{equation}
\begin{aligned}
a=-\frac{1}{8 \pi M_{\mathrm{th}}} T\left(M^2_{\mathrm{th}}\right),
\end{aligned}
\end{equation}
where $M_{\mathrm{th}}$ is the $DK$ threshold, and $T$ is the scattering amplitude in Eq. (\ref{eq:BSE}).
In our calculation, we have $a=(-0.12+0.55i)$ fm.
Its absolute value is close to the theoretical result in Ref. \cite{Guo:2009ct}, but larger than the lattice QCD calculation \cite{Liu:2012zya}, while smaller than the data extracted using the K-matrix model in the LHCb experiment \cite{LHCb:2024iuo}.

\begin{figure}[htbp]
\centering
\includegraphics[width=0.8\linewidth,trim=0 0 0 0,clip]{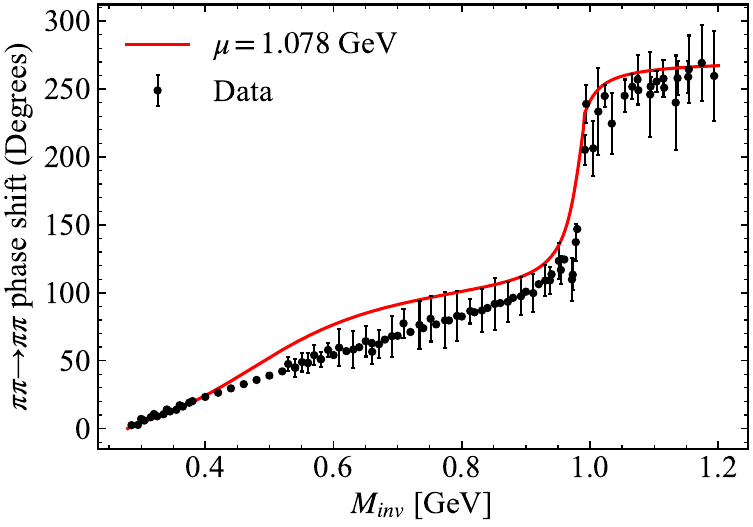} 
\caption{The $\pi\pi\rightarrow\pi\pi$ phase shift of $I(J)=0(0)$ obtained with the parameters used in this work. The red curve is our result with the regularization scale $\mu=1.078~\text{GeV}$, and the data without errors are taken from Refs. \cite{Froggatt:1977hu,NA482:2007xvj}, while the ones with errors are the averaged data taken from Refs. \cite{ochs74,refpipiphase,kaminski97zpc}, as used in Refs. \cite{Oller:1998zr,Guo:2012yt}.}
\label{fig:pipiPhaseShift}
\end{figure}

On the other hand, in the lower row of Fig. \ref{fig:TXX}, we can see that the $f_{0}(500)$ resonance appears in the amplitude from $\pi\pi$ to $\pi\pi$ channel, while the $f_{0}(980)$ resonance appears in all three amplitudes.
The pole position corresponding to the $f_{0}(500)$ resonance is $\sqrt{s_{p}}=(0.471-0.181i)$ GeV when taking $\mu=1.078$ GeV.
Its mass is as same as the experiment data, while its width $0.362$ GeV is more than $60\%$ larger than the experimental value $0.224$ GeV \cite{LHCb:2024iuo}.
Therefore, it will have a significant effect on the properties of the $T_{c\bar{s}}$ state.
Actually, the $f_{0}(980)$ state can also be reproduced in the $S$-wave scattering amplitude of the $\pi\pi$ coupled channel system, but it is not in the energy range of our interest, and thus can be ignored.
In addition, the parameters obtained from the fitting in this paper can also well describe the phase shift of $\pi\pi\rightarrow\pi\pi$ with $I(J)=0(0)$, as shown in Fig. \ref{fig:pipiPhaseShift}.

Based on the above results, we are confident that this new $T_{c\bar{s}}$ state is dynamically generated in the $S$-wave interactions of the $D_{s}\pi$ and $DK$ coupled channels.
It is similar to the $T_{c\bar{s}}(2900)$ state for the $D_{s}^{*}\rho$ system and the $X(5568)$ state for the $B_{s}\pi$ system in Refs. \cite{Albaladejo:2016eps,Molina:2022jcd,Duan:2023lcj}.
A difference from the Ref. \cite{Molina:2022jcd} is that, in this work no pole is found, and the $T_{c\bar{s}}(2900)$ is classified as a threshold effect, where there is also a cusp structure in the amplitude.
It should be emphasized that although there are clear cusp structures in the amplitudes in above row of Fig. \ref{fig:TXX}, the calculated invariant mass distributions in below row of Fig. \ref{fig:Figs123} do not have this characteristic.
Further experimental measurements are necessary to further determine the properties of this new $T_{c\bar{s}}$ state.
We suggest experimental colleagues to further explore its properties in bottom mesons decays.

\section{Summary}\label{sec:Summary}

Over the past two decades, the exotic hadronic states has consistently been a hot topic. Their mass spectrum, structures, and decay properties are crucial for the understandings of the behaviours of quantum chromodynamics in non-perturbative energy regimes.
Both theoretical and experimental efforts have been made to search and understand them.
The quark content of the $T_{c\bar{s}}$ state discovered by the LHCb Collaboration in the $D_{s1}(2460)^{+} \rightarrow D_{s}^{+}\pi^{+}\pi^{-}$ reaction shows that it is a candidate for an exotic state.
In this work, we investigate the properties of this state and its contribution, together with the $f_{0}(500)$ resonance, in the $D_{s1}(2460)^{+} \rightarrow D_{s}^{+}\pi^{+}\pi^{-}$ decay.

We calculate the amplitude based on the final state interaction within the chiral unitary approach.
The fitted $D_{s}^{+}\pi^{+}$ and $\pi^{+}\pi^{-}$ invariant mass distributions agree well with experimental data.
We find that without the $D_{s}\pi$ coupled channel interaction, considering only the tree-level and the $\pi\pi$ coupled channel interaction cannot reproduce the experimental data.
The two peaks in $D_{s}^{+}\pi^{+}$ invariant mass distribution are mainly contributed by the doubly charged $T_{c\bar{s}}^{++}$ and its isospin partner $T_{c\bar{s}}^{0}$.
Besides, the first peak in $\pi^{+}\pi^{-}$ invariant mass distribution is mainly contributed from the $f_{0}(500)$ resonance, and the second one is contributed by the reflections of the $T_{c\bar{s}}^{++}$ and $T_{c\bar{s}}^{0}$ states.
Furthermore, distinct cusp structures of $T_{c\bar{s}}$ near the $DK$ threshold can be observed in the $D_{s}\pi$ coupled channel $S$-wave scattering amplitudes.
Its corresponding pole is $\sqrt{s_{p}}=(2.394-0.057i)$ GeV on the second Riemann sheet, of which the real part depends on the theoretical free parameter $\mu$, but the imaginary part close to $0.057$ GeV remains almost unchanged.
We think that the $f_{0}(500)$ and the new $T_{c\bar{s}}$ resonances are the molecules of $\pi\pi$ and $DK$, respectively.
More experimental data are imperative to further investigate the properties of the $T_{c\bar{s}}$ states in detail, and we expect experimental colleagues can continue to pay attentions to these states.

\section*{Acknowledgements}

We would like to thank Prof. Eulogio Oset for his valuable comments.
This work is supported by the Natural Science Special Research Foundation of Guizhou University Grant No. 2024028 (Z.-Y. W.), and the National Natural Science Foundation of China under Grants No. 12405098.

\addcontentsline{toc}{section}{References}

\end{document}